\documentstyle[aaspp4,epsfig,psfig]{article}
\def \calb{{\cal {B}}}

\def \vF  {{\bf {F}}}
\def \F {\vF_{\Delta}}
\def \vA1  {\bf {A}_1}

\def \vA  {\bf {A}}

\def \vri {{\bf {r}}_i}
\def \vr  {{\bf {r}}}
\def \vpi {{\bf {p}}_i}
\def \vp {{\bf {p}}}
\def \vvi {{\bf {v}}_i}
\def \vv {{\bf {v}}}
\def \vx {{\bf {x}}}
\def \vecX {{\bf {X}}}

\def \vw  {\bf {w}}
\def \grad {{\bf {\nabla}}}
\def \vecR  {{\bf {R}}}
\def \vecV  {{\bf {V}}}

\def \fop {{f^{\rm {op}}}}
\def \mr {{<r^2>^{1/2}}}
\def \vir {{<r^2>}}
\def \si {{<v^2>^{1/2}}}
\def\ltord{\hbox{$\;\raise.4ex\hbox{$<$}\kern-.75em\lower.7ex\hbox{$\sim$}
                           \;$}}
\def\gtord{\hbox{$\;\raise.4ex\hbox{$>$}\kern-.75em\lower.7ex\hbox{$\sim$}
         \;$}}

\begin{document}

\title{Dynamical Friction and The Evolution of Satellites  in Virialized
 Halos: The Theory of Linear Response}
%\centerline {and}

%\title {Dynamical Friction of Satellites inside Virialized Halos}

\author{Monica Colpi\altaffilmark{1}, Lucio Mayer\altaffilmark{1}, \&
Fabio Governato\altaffilmark{2}}

\altaffiltext{1}{Dipartimento di Fisica, Universit\`a Degli Studi di Milano, 
Milano, Via Celoria 16, I--20133 Milano, Italy}
\altaffiltext{2}{Osservatorio Astronomico di Brera, Merate,
via Bianchi 46, I--23807 Merate (LC) - Italy} 

\begin{abstract}
The evolution of a small satellite inside a more massive truncated
isothermal spherical halo is studied using both the Theory of Linear
Response for dynamical friction and  $N$-Body simulations.  The
analytical approach includes the effects of the gravitational wake, 
of the tidal deformation and the shift of the barycenter of the primary,
so unifying the local versus global interpretation of dynamical
friction.
The $N$-Body simulations followed the evolution of both rigid and
live satellites within larger systems. Sizes, masses, orbital energies
and eccentricities are chosen as expected in hierarchical clustering
models for the formation of structures.
Results from this coupled approach are applicable
 to a vast range of astrophysical problems,
from galaxies in galaxy clusters to small satellites of individual galaxies.
The main contribution to the drag results from the gravitational pull
of the overdensity region trailing the satellite's path since  the stellar
response to the external perturbation remains correlated over  a time
shorter than the typical orbital period. 
The analytical approach and the $N$-Body experiments demonstrate that
there is no significant circularization of the orbits and that 
the dynamical friction time scale is  weakly
dependent on the circularity $\varepsilon$.
While the theory  and the $N$-Body simulations 
give a complete description of the orbital decay of satellites
a good fitting formula for the orbital decay time is:
$$\tau_{DF}=1.2 {
J_{cir}r_{cir}
\over[GM_{sat}/{\rm{e}}]\ln(M_{halo}/M_{sat})}\,\varepsilon^{0.4} 
$$
where $J_{cir}$ and $r_{cir}$ are, respectively, the initial
orbital angular momentum and the radius of the circular orbit with the
same energy of the actual orbit. Tidal stripping can reduce the
satellite's mass by 60\% after the first pericentric passage 
increasing the orbital decay time.  The ``e"  factor keeps that effect
into
account and should be removed in the simplified case of  rigid satellites.
In cosmologically relevant situations our model gives orbital decay 
times larger by a factor of $2$ with respect to most previous estimates.

For peripheral orbits where the apocenter is larger than the virial radius
of the primary decay, the global tidal field and the shift of the barycenter
become important.
In this case $\tau_{DF}$ needs to be further
increased by at least $\simeq 50\%$.
The final fate of a satellite is determined by its robustness against
the effect of tides.  While low density satellites are disrupted over
a time comparable to the decay time of their rigid counterparts,
satellites with small cores can survive up to an Hubble time
within the primary, notwithstanding the initial choice of orbital
parameters. 
Dwarf spheroidal satellites of the Milky Way, like Sagittarius A and  
Fornax, have already suffered
mass stripping and, with their present masses, the sinking times exceed 10 Gyr
even if they are on very eccentric orbits.
 
\end{abstract}

\section{Introduction}

Dynamical friction is a fundamental physical process that drives the
evolution of most cosmological structures, from satellites in
galaxies, to galaxies in large clusters. 
The satellites or the galaxies   can decay toward the center of the
halos as friction causes the loss of orbital energy and 
angular momentum. 
This braking force is a force of back-reaction resulting from
the global distortion of the stellar density field induced in the
primary by the satellite's gravity.  

The theory of linear response (TLR) is an ideal tool for studying the
dynamics of sinking satellites in spherical  halos.  Recently explored
by Colpi \& Pallavicini (1998, CP in the following), this formalism is
alternative to other perturbative techniques developed to overcome the
limits of Chandrasekhar's theory of dynamical friction (Chandrasekhar 
1943), that restricts to infinite, uniform, non selfgravitating
stellar backgrounds.
In TLR, the force can either be related to the density changes, or be
viewed as a direct manifestation of the fluctuation-dissipation
theorem (CP; Nelson \& Tremaine 1997; Bekenstein \& Maoz 1992;
Kandrup 1981). In the last interpretation, the fluctuations of the
microscopic two body force (between the satellite and the particles)
add to give a nonvanishing drag.  The force is the result of a time
integral keeping memory of the actual dynamics of the satellite and of
the dynamics of the stars as derived from their Hamiltonian.  The
stellar reaction to the external perturbation and the amount of
dissipation that follows depend closely on the (self)-correlation
properties of the underlying stellar field (Colpi 1998, C98
hereafter).

TLR has been successful in describing the decay of satellites
spiraling outside their primary spherical halos, and in describing
their deceleration in shortlived penetrating encounters.  In the first
case, global tidal deformations excited during orbital motion are
responsible to the loss of stability of the satellite's orbit (C98).
In flybys (CP), the deceleration was viewed
as a superposition of two effects: the gravitational pull
from the overdensity region that forms just behind the satellite's
trail and the brake from extended tides.
 
Cosmological simulations aimed at studying the building up of cosmic
structures (Ghigna et al. 1998; Tormen et al. 1998; 
Kravtsov \& Klypin  1998) show
that the majority of the satellite's (or substructure) orbits have
rather large eccentricities, and apocenters that rarely exceed twice
the virial radius of the primary halo (defined as the radius where 
the average density inside the halo is 200 times the critical one). 
In this work we consider the fate of satellites placed on 
cosmologically relevant orbits, using TLR and $N$-Body simulations
as a tool for exploring the decay in a self-gravitating spherical halo.

TLR is a major advance relative to alternative 
analytical studies (Weinberg 1986; Weinberg 1989; S\'eguin \& Dupraz 1994)  
as it embraces in a relatively simple way all aspects 
of the gravitational interaction of
the satellites with the collisionless background: the wake and the tides in a
self-gravitating system, and the shift of the
stellar center of mass.
We then explore the dependence of the sinking times as a function of 
the orbital parameters and the possible role played by 
mass stripping on their  evolution.
A previous analysis 
was carried on by Lacey \& Cole (1993) in their
semi--analytical
treatment of the merging of cosmic structures.
The aim was to clarify the role of dynamical friction in determining
the merging rate of the luminous part of galaxies as opposed to
the merging rate of their dark matter halos.
A dependence of the decay time on eccentricity was introduced 
by simply fitting the decay time curve obtained using
Chandrasekhar's formula for a point-like satellite and for
a fixed orbital energy (the orbital energy was such that a circular
orbit had a radius equal to the virial radius of the primary halo).
The authors found that satellites on nearly-circular orbits decay
on a timescale which is almost $3$ times longer than that of
satellites on very eccentric orbits.
However satellites accreting on larger halos are usually on more
tightly bound orbits according to cosmological simulations, i.e. even
radial orbits have apocenters close to the virial radius
of the main halo (in the Lacey \& Cole 
case the apocenter of radial orbits was well outside the virial
radius).  A more appropriate choice of the orbital parameters is thus
needed to determine to what extent the sinking times depend on the
eccentricity of the orbits.  Moreover, dark matter halos of individual
satellites would not dissolve immediately: Navarro et al. (1995),
using $N$-Body/SPH simulations, have shown that discarding the dark
matter mass can lead to an overestimate of the sinking times of the
baryonic cores.
It is thus evident that a
complete description of the orbital evolution of satellites has
to take into account the effect of tidal stripping on the decay
times of satellites.
It also necessary to find out if the orbits circularize during
the decay, this being another long-standing issue.
If orbits do circularize, we expect the initial distribution of orbital
eccentricities of the satellites to be substantially altered,
and this could prolong their lifetimes in the primary halos, provided
that a substantial mismatch between decay times on eccentric and
circular orbits does really exist. 

Recently, van den Bosch, Lewis, Lake \& Stadel (1999, vBLLS hereafter)
carried on a series of $N$-Body simulations of
satellites placed inside a nonsingular tidally limited spherical halo.
They did not include tidal stripping treating the satellites as rigid
spheres.
To study the orbital evolution of satellites in a fully
self-consistent way and to complement the results obtained with TLR, we
have performed a number of high resolution $N$-Body simulations with
the parallel binary treecode PKDGRAV (Stadel \& Quinn 1998; Dikaiakos \&
Stadel 1996;  vBLLS) for
both rigid and deformable satellites.

The outline of the paper is as follows: in $\S2$ we give a brief
description of the theory of linear response for dynamical friction
setting the framework for the calculation carried in $\S3$. In $\S4$
we illustrate the results of our semianalytical study and compare them
with vBLLS.  In $\S5$ we explore the self-correlation properties of
the equilibrium stellar system in terms of the correlation time scale
and study the fading of the density wake in a uniform
infinite medium. In $\S6$ we extend our analysis
to the important case of non-rigid
satellites. $\S7$ contains our conclusions.

\section {The theory of Linear Response for Dynamical Friction}

A satellite bound to a primary galaxy experiences in its motion a
dissipative force that results from the collective response of the 
background to its perturbation.
In TLR, the response depends just on the properties of the underlying
matter field in its unperturbed state: The (self)-correlations existing
among the particles 
ultimately leads to energy dissipation.

Under the hypothesis that the $N-$ stars (of mass $m$) or dark matter
particles are in virial spherical equilibrium, the drag force $\F$ on a
satellite described as a point like object of mass $M$ 
reads 
\begin{eqnarray} 
\F(t)= & &
GMm\sum_{i=1}^N\,\,
\int_{t_0}^{t}ds 
\nonumber \\ & &
\int d\Gamma\,\,\left
[\nabla_{\vpi(s)}f_0 {\bf {\cdot}} 
{\vecR(s)-\vri(s)\over \vert
\vecR(s)-\vri(s)\vert^3}\right ] 
\left [ GMm\sum_j^N{\vecR(t)-\vr_j(t)\over \vert
\vecR(t)-\vr_j(t)\vert^{3}}\right ]
\end{eqnarray} 
where 
$d\Gamma$ is the elementary volume  in the $6N$ dimensional phase space
($\Gamma$) of the stars in the galaxy, and $f_0$ the
$N$-point equilibrium distribution function
(we drop the subscript
$sat$ hereafter to follow closely the notation of CP
and denote the total halo mass $M_{halo}$ as $Nm$). 
The drag on $M$ is  a consequence of a memory effect that
develops with time and involves a suitable phase space average of the
microscopic two-body force. 
It requires the knowledge of the dynamics of
the $N$ stars $(\vr_i(s),\vp_i(s)),$ as determined by the unperturbed 
Hamiltonian, over the whole interaction time from $t_0$ (
when the perturbation is turned on) to  the current time $t$.
The distribution function $f_0$ incorporates  the 
properties of the system in virial equilibrium.

Dark matter halos  in virial equilibrium  can be
regarded as an assembly of collisionless particles subject to   
a mean field potential $\Psi_0$  that can be computed  
solving simultaneously the Poisson and Boltzmann equation.
The distribution function can thus  be written in terms of the
one-particle phase
space density $\fop(\vr,\vp).$ 
Under this hypothesis, and due to the statistical independence of
the particles,
all cross correlation terms cancel identically in the limit of $N\gg 1$.
Only the self-correlation properties of the collisionless background  survive to yield
\begin{eqnarray}
\F(t)= & &
[GM]^2\,Nm^2\,\int_{t_0}^t ds
\,\int d_3\vr \, d_3\vp 
\nonumber \\ & &
\left \{ \nabla_{\vp(s)}\fop\cdot \left [
{\vecR(s)-\vr(s)\over \vert \vecR(s)-\vr(s)\vert^{3}}-
\int \,d_3\vr' n_0(r') 
{\vecR(s)-\vr'\over \vert \vecR(s)-\vr'\vert^{3}}\right ]\right \}
{\vecR(t)-\vr(t)\over \vert \vecR(t)-\vr(t)\vert^{3}}
\end{eqnarray}

The new term appearing in brackets 
(involving the equilibrium background  density $n_0(r)$)  
represents, at
a given time $s,$ the mean force acting on $M$ resulting  from
the  system as a whole: It accounts for
the shift of the center of mass of the galaxy during the encounter.
The recoil of the halo (due to linear momentum conservation) is a coherent shift of all
the orbits of the background particles
, giving origin to  a global correlation among them.
TLR, as is formulated, can account for that shift naturally and permits use of
the one-particle distribution function $\fop$ for the system in virial equilibrium  
(see CP and C98).  Thus, $\F$ in the form of equation (2) is  the force as
measured  in the
non-inertial reference 
frame comoving with the halo's center of mass. 

In the context of the fluctuation-dissipation theorem, the braking force
can be seen as an integral over  time  of the  
correlation function of a fluctuating component of the microscopic
 force  
\begin{equation}
F^a_{\Delta}(t)
\equiv \int_{t_0}^t \,ds \,K^a(t-s)=\int_{t_0}^t \,ds \int d_3\vr \,d_3\vp  
\grad^b_{\vp(s)}\fop \,\,T^{ba}
\end{equation}
where the self-correlation 
tensor reads
\begin{equation}
T^{ba}\equiv [GM]^2Nm^2 \left [ {R^b(s)-r^b(s)\over \vert
\vecR(s)-\vr(s)\vert^{3}}\,\,
- \int \,d_3\vr' n_0(r') 
{R^b(s)-r^{'b}\over \vert \vecR(s)-\vr'\vert^{3}}\right ]
{R^a(t)-r^a(t)\over \vert \vecR(t)-\vr(t)\vert ^3}.
\end{equation}
The correlation function $K^a(t-s)$  
introduces a  
time scale  $\tau^*$ characterizing the rise time of the
force $F_\Delta ^a(t):$ It is the scale over
which the stars  redistribute  the satellite's orbital energy into the internal
degrees of  freedom of the system.

The interpretation of $\F$ in terms of a global 
time dependent density deformation is possible also within TLR, noting that
equation (2) can be written formally as 
\begin{equation}
\F=
-GMNm\int d_3\vr ~\Delta n(\vr,t) ~{\vecR(t)-\vr\over
\vert \vecR(t)-\vr\vert^{3}}
\end{equation} 
where the function $\Delta n(\vr,t)$ maps  the 
response, i.e., the time dependent  changes in the density field
$n_0(r)+\Delta n(\vr,t)$ resulting from the superposition (memory)  of
disturbances
created
by the satellite over the entire evolution; 
the function $\Delta n(\vr,t)$ can be derived comparing equation (5) with
(2) (see also CP for details).

Equation (2) applies when the
interaction potential between $M$ and the stars is weak relative to the
mean field potential $\Psi_0$ of the equilibrium system (when in isolation).
This is the reason way only the properties of the halo in virial
equilibrium are requested to evaluate $\F$.
As a consequence of that $\F$ is accurate to
second order in the coupling constant $G.$ Higher orders terms would
describe the self-gravity of the response, i.e., the modification
in the self interaction potential 
due to the external perturbation driven by $M$.
Equation (2) can describe the sinking of satellites moving on arbitrary orbits, even
outside the primary halo. Previous semianalytical studies focussed on purely circular
orbits (Weinberg 1986) to explore the role of resonances and on almost
radial orbits to
explore the transient nature of the interaction (S\'eguin \& Dupraz
1994).

\section {TLR: the Force of Back-reaction in a Spherically
Symmetric Galactic Halo} 

In a nonuniform collisionless background the back-reaction force on 
$M$ results, in the high speed limit,  
from the  combined action of  a global tidal response 
related to the density gradients (absent in an infinite uniform medium) 
and  from the development
of an extended wake forming behind the
satellite 's path that contributes mostly to its deceleration.
The force acquires a component along $\vecR$ as symmetry around $\vecV$ is lost,
the underlying system being non homogeneous.

To estimate the drag in the domain where the satellite's velocity $\vecV$
(determined primarily by the mean field potential $\Psi_0$ of the unperturbed
background) 
maintains comparable
to the background velocity dispersion, we avoid to separate out the  
tidal and
frictional contributions being aspects of a unique process. 
Exploiting the time independence of the
distribution function $\fop$ and of the phase-space volume
$d_3\vr\, d_3\vv$ ($\fop$ hereafter will
be considered as a function of $\vr$ and $\vv=\vp/m$ and is normalized
accordingly), the
drag force
(eq. [2])
can be equivalently written as

\begin{eqnarray} 
\F= & &
[GM]^2Nm\int_{t_0}^t\, ds \int d_3\vr \,d_3\vv
\,\, {\grad_{\vv}}\fop(\vr,\vv) \, {\bf {\cdot}} 
\nonumber \\  & &
\left [\grad_{\vecR(s)} \phi(\vert \vecR(s)-\vr\vert) - 
\int d_3\vr'\, n_0(r') \grad_{\vecR(s)} \phi(\vert\vecR(s)-\vr'\vert)\right ]
\grad_{\vecR(t)}\phi(\vert\vecR(t)-\vr(t-s)\vert)
\end{eqnarray}
where $\phi$ is proportional to the newtonian gravitational potential 
\begin{equation}
\phi (\vert \vecR-\vr\vert) \equiv -{1\over \vert \vecR-\vr\vert}.
\end{equation}
In equation (6), $\vecR$ denotes the satellite position vector relative to
the halo's center of mass, and is computed self-consistently following the actual
dynamics of the
satellite (that now acquires the reduced mass $\mu$).

Because of the difficulty of including the dynamics of the stars as 
determined by the
unperturbed Hamiltonian we are led to approximate their motion as
linear
giving 
\begin{equation}
\vr(t-s)=\vr+(t-s)\vv.
\end{equation} 
We will
compare our model  with $N$-Body simulations (described in $\S4$)  to
test indirectly the validity of such an approximation.  
In neglecting the acceleration of
the stars, i.e., their ``curvature'', during the interaction of the
satellite we introduce a
simplification which will prove to be satisfactory.  

The shortcoming of TLR is its
inability to describe short distance encounters as it is derived from a 
linear analysis expanded to first order in the perturbation.
For a pointlike satellite moving in an infinite  uniform medium, these
encounters    
lead to a minimum impact parameter which is  determined
uniquely by  $V$ and the background velocity dispersion $\sigma$.
Satellites have finite size and as in $N$-Body
simulations the short-distance two-body interaction $\phi$ is  smoothed
introducing in
the microscopic gravitational  potential a softening length $\epsilon$. 
The necessity of
a closer comparison
with numerical simulations led us to consider the spline kernel potential $\phi_{sp}$
(Hernquist \& Katz 1989) as interaction potential 
between the satellite and the stars reducing to the newtonian form (eq. [7]) 
at $2\epsilon$. 
%Hence, we derived a semianalytical  expression of
%$\F$  substituting $\phi$ with 
%$\phi_{sp}.$ 
The introduction of the softening
length in the computation of the force accounts for the finite size of $M$
permitting an unambiguous comparison with the numerical simulations
by vBLLS.
The drag force depends on the 
response of the stars  
and, in turn, on the characteristics of their  
equilibrium  state which is described below.

Dark halos are often modelled
as truncated
non-singular isothermal spheres with a core
(vBLLS; Hernquist 1993): accordingly, their density
profile 
\begin{equation}
n_0(r)= {1\over 2\pi^{3/2}\,g\,r_t}  {\exp(-r^2/r_t^2)\over (r^2+r_c^2)}
\end{equation}
declines  exponentially at radii exceeding  the tidal (or truncation)
radius  $r_t.$ 
The homogenous core of radius $r_c$ is surrounded by a region
where $n_0(r)\propto r^{-2},$ as in a  singular
isothermal sphere. The constant 
\begin{equation}
g=1-\pi^{1/2}\left ({r_c\over r_t}\right )\,\exp(r_c^2/r_t^2)\,\left [ 
1-{\rm{erf}} \left ({
r_c\over r_t}\right )\right ]
\end{equation}
is introduced to guarantee that  $\int d_3\vr n_0(r)=1.$

The one-dimensional background velocity dispersion $\sigma$ is computed
according
to the second-order Jeans equation
\begin{equation}
\sigma^2(r)\equiv {1\over n_0(r)}\int_r^{\infty}dr'n_0(r'){4\pi
G\over r'^2}\int_0^{r'}
dr'' \, (r'')^2n_0(r'')
\end{equation}
and is a local function of $r.$ 
For $r_c\to 0$
\begin{equation}
\sigma^2(r)={GNm\over g\,r_t}\left ({r\over r_t}\right )^2
\exp(r^2/r_t^2)
\int_{(r/r_t)}^{\infty} dx \,\exp(-x^2)\, x^{-4} \,{\rm{erf}}(x)
\end{equation}
giving  
$\sigma^2\simeq GNm/(g \,r_t \sqrt{\pi}),$
at $r\ll r_t.$ 
The back-reaction force on $M$  is derived under the hypothesis
that the one-particle distribution function is isotropic and Gaussian
in the velocity space, with $\sigma^2$ computed according to equation
(12):
\begin{equation}
\fop(\vr,\vv)=n_0(r)\left ( {1 \over 2 \pi\sigma^2 }\right )^{3/2}
\exp\left (-v^2/(2\sigma^2)\right ).
\end{equation}
This choice is dictated not only by simplicity arguments but by the fact
that   collisionless systems with these characteristics are found to be
nearly in equilibrium (Hernquist 1993; vBLLS) and hence 
are viable for describing the unperturbed system equilibrium state
required by TLR.
Given $\fop,$  we compute $\F$ from equation (6).
Not all multiple integrals of equation (6) can be carried out analytically; 
only those over the velocity
phase space are evaluated and the
complex expression of the drag is reported in the
Appendix.

The
evolution of a satellite, of reduced mass $\mu=MNm/(M+Nm),$  is followed  
solving for the equations of motion, in the reference frame comoving
with
the center of mass of the primary halo:
\begin{equation}
\mu {d^2\vecR(t)\over dt^2}= - {GNm}{\vecR(t)\over  \vert \vecR(t)\vert ^3
}\int_{r'<R(t)} \, d_3\vr'\, n_0(r')
+\F
\end{equation}

%The force resulting from the mean unperturbed ### potential is
%computed through a series expansion and the drag force is determined
%solving for a set of multiple integrals.
%The satellite is set initially along  orbits of given energy and eccentricity.

The mass ratio $M/Nm$ and the cusp $\epsilon,$ entering the effective
 potential $\phi_{sp},$ are the only parameters of the model.

\section {The Sinking of the Rigid Satellites }

In this section we explore the evolution of satellite orbits comparing
results obtained using TLR with our $N$-Body runs and, where possible,
with those of vBLLS.  As in vBLLS, the primary system, scaled to the
Milky Way's halo, is a spherical isothermal halo with a mass of
$10^{12}M_{\odot}$, a tidal radius $r_t$ of 200 kpc and a core radius
$r_c=r_t/50$. In the $N$-Body simulations the primary halo has $10^4$
particles: it was first evolved in isolation for $10$ Gyr and the
stability of the density profile was verified.  The satellite is
fifty times lighter than the primary ($M=Nm/50$).  Its mass
distribution is described by a rigid spline softened potential with
a length scale of 3.4 kpc, comparable to the effective radius of the
Large Magellanic Cloud ($\epsilon=0.0172\,r_t$).  With these choices
the time unit  $T_0=[GNm/r_t^3]^{(-1/2)}$ is of $1.34$ Gyr.

\subsection {The Dynamical Friction Decay Time and the Evolution of $e$ for
Rigid Satellites }

Cosmological simulations have shown  that in the hierarchical
clustering scenario most of  satellite's
orbits have pericenter varying between $0.2<r_{peri}/r_{t}<0.5 $ and
apocenter $r_{apo}<2r_t$ (Ghigna et al. 1998).  More loosely bound
orbits are unlikely as their apocenter can exceed the turnaround
radius (of about 2$r_t$) of the major overdensity that produced the
primary halo.  Moreover, orbits are found to be quite eccentric on
average, with a typical apocenter to pericenter distance ratio $ \sim
6-8$ corresponding to eccentricities between 0.6 and 0.8.  Below we
focus attention on orbits with reference circular radii $r_{cir}$
(determining the initial energy) in the range $0.5 \leq r_{cir}/r_t \leq 1$
(to fulfill the above inequalities). We are
  thus able to study the dependence
of $\tau_{DF}$ both on eccentricity and orbital energy.

Figure 1 shows the dynamical evolution of the satellite set on bound
orbits having initially equal energy but different eccentricity:
$e=0.8$ (top left panel),$0.6,0.3,$ and 0 (lower right panel)
respectively.  The radius of the circular orbit at the onset of evolution 
is $r_{cir}/r_t=0.5$
as in vBLLS: the
selected runs coincide with models 3,4,5,6 (we refer to the table of
content 1 of vBLLS).  To characterize the decay and quantify the
results, we report in Figure 2 the angular momentum as a function of
time (in Gyr); dots are the results of the $N$-Body simulations
carried out by vBLLS.  The agreement between theory and 
$N$-Body simulations is excellent to a few \%.

The analysis of more loosely bound orbits has been carried out using
both TLR (for $r_{cir}/r_t=0.8,1$) and $N$-Body simulations
($r_{cir}/r_t=1$).  The results show once more an excellent agreement
between theory and simulations (Fig. 3). Minor differences might be
caused by the limited resolution of our $N$-Body galaxy model: in
particular, the potential sampling may be exceedingly poor in the
outer region of the halo, where the satellites now spends a lot more
time.  We have tested this hypothesis performing a high resolution
simulation with $10^5$ particles: in this case there is a closer
agreement (Fig. 3), proving the potential of the theory with respect
to costly $N$-Body simulations.

To evaluate the decay time of a satellite moving 
within  a singular isothermal sphere  Lacey \& Cole (1993)  
proposed the following general expression  
to incorporate the dependence of the sinking time on the 
initial eccentricity and orbital energy:
\begin{equation}
\tau_{DF}=T_c~\varepsilon^{0.78}\equiv 1.17~{r^2_{cir}V_{cir}\over
GM\ln(Nm/M)}\varepsilon^{0.78}
\end{equation}
In the formula $V_{cir}$ is the circular velocity of the satellite,
$r_{cir}$ is the radius of the circular orbit having the same
energy of the actual orbit and the circularity 
$\varepsilon\equiv J(E)/J_{cir}(E)$
is the ratio between the orbital angular momentum and that 
of the circular orbit having the same energy $E$.
Lacey \& Cole suggested a value of $\alpha=0.78$ for the dependence
on eccentricity.
However,  as already noticed by vBLLS, the decay time $\tau_{DF}$ depends more weakly on
the initial eccentricity and for the case  $r_{cir}/r_t=0.5$
vBLLS proposed a best fit  of the form  $\tau_{DF} \propto
\varepsilon^{\alpha}$ whose exponent $\alpha$=0.53.

Both TLR and our $N$-Body simulations show that $\alpha$ depends on the
energy
of the orbit and that the value of the scale $T_c$ deviates slightly from the one
inferred using Chandrasekhar's formula.
For the cosmologically relevant orbits, the
TLR approach (supported by our set of $N$-Body simulations), gives a 
$\tau_{DF} \propto \varepsilon^{\alpha}$ whose exponent $\alpha$=0.4.

For the case of orbits outside the halo (C98),
satellites on wider orbits not only have longer decay times but
the mismatch between very eccentric and nearly circular orbits  
becomes increasingly larger with decreasing orbital energy.
A continuity in the behaviour of $\tau_{DF}$ should therefore
exist when moving from peripheral orbits to internal ones.
Figure 4 shows $\tau_{DF}(\varepsilon, r_{cir}),$ computed using TLR, for
$r_{cir}/r_t=0.5,0.8,$ and 1.
Dots refers to eccentricities $0(\varepsilon=1),~0.3,0.6,0.8,0.85$
respectively.  As mentioned in the above paragraph, the best fit gives a slope
$\alpha$ of $0.4$ for $r_{cir}/r_t=0.5.$   
We find that for the typical eccentric orbits (Fig. 4)
occurring in current structure formation scenarios 
the dynamical friction time scale is longer by  a factor of 1.5-2
relative to previous estimates (Lacey \& Cole 1993), a result that may
affect significantly the statistics of the satellites in galaxy halos and
of galaxies in galaxy clusters. 

One still open problem is if orbits tend to circularize under the
effect of dynamical friction.  This could be true if halos lose energy
faster than angular momentum.  It has been recently shown (Ghigna et
al. 1998) that satellites inside larger halos have a distribution of
orbital eccentricities quite indistinguishable from that of the
diffused dark matter component.  This strongly suggests that
orbits of
satellites do {\it not} evolve significantly under the effects of
tidal stripping and dynamical friction.
We can now support this numerical result within TLR 
showing that  bound  orbits 
are not subject to  any significant circularization, as illustrated  in
Figure 5; the result holds also when exploring the evolution of
live satellites. 
Only when the
satellite   happens to fall from the very far reaches of the halo
(in grazing encounters), TLR
predicts some degree  of circularization (C98), 
consistent with numerical simulations  by Bontekoe \& van
Albada (1987).

\section{ The Self-Correlation Properties of the Fluctuating Microscopic
Force}

\subsection {The Self-Correlation Time}

An important results of this study is that two independent
calculations, i.e., a semianalytic theory and a set of $N$-Body
simulations give equivalent results.  The quite stringent accordance
between the two approaches confirm the applicability of TLR in
describing the sinking of satellites in spherical nonhomogeneous
halos, under the neglect of the actual stellar dynamics.

A question thus rises naturally:
What is the 
role played by the self-gravity of the background in determining the extent of
the drag ?
Equation (6) contains many aspects of the
self-gravity of the 
collisionless background (in its unperturbed state): 
(i) the equilibrium dynamics  that 
establish the strength of the self-correlation properties of the system, 
(ii) the density profile  $n_0(r)$,  (iii) the virial relation
that links 
$n_0(r)$  to the dispersion velocity $\sigma(r)$
and ultimately, (iv) the shift of the system barycenter.
The braking torque depends on all these
quantities that are interrelated.

The dynamics of the stars in the unperturbed potential $\Psi_0$ is
expected to be important in the determination of $\F$ if
their response to the external perturbation remains
correlated   for a time $\tau^*$ longer than the typical radial period
$T_r=2\pi r_{cir}/V_{cir}$
(where $V_{cir}$ is the circular velocity 
in the primary halo), which is of a few time units, for the mean field potential
$\Psi_0$ generated by the density distribution of equation (9).
$\tau^*$ can be 
estimated using  equations (3) and (4):
Figure 6 shows the cumulative function 
\begin{equation}
I^a(s)\equiv \int_{t_0} ^s \, ds'\, K^a(t-s')
\end{equation}
at four selected times ($t=2,3,4,6,8$ time units), during orbital
evolution, for
$e=0.6$ and $r_{cir}=0.8 \,r_t$. We find that 
the rise of the correlation function $K^a$ is rather rapid
as it occurs over a time $\tau^*$  which is only $\simeq 1\,T_0$, i.e., a
fraction of the typical radial time $T_r$. 
On this scale $\tau^*,$ stars thus follow a dynamics that can be approximated  as
free: This verifies the internal consistency of our calculation
and explain the equivalence between  the analytical and numerical calculation.

A finite $\tau^*$ does not necessarely imply $\tau^*<T_r.$ 
The self-correlation time scale can exceed  $T_r:$  
Memory can be maintained over many orbital periods and resonant transfer  of
energy can
accelerate dramatically the decay. 
As found in C98, a satellite moving outside a spherical halo experiences a drag  
resulting from the global tidal deformations excited by the satellite itself. 
The drag force was found, generally, to be
\begin{equation}
F^a_{\Delta}=-[GM]^2{Nm\over \sigma^2}O^{abc}(t)\int_{t_{0}}^t\,ds
\calb(t-s)Q^{bc}(s)
\end{equation}
where the tensors $Q$ and $O$ represent the quadrupolar and octupolar 
terms leading 
the multipole expansion of the interaction potential 
(Prugniel \& Combes 1992 recognized  the importance of these terms in
their numerical simulations). 
In equation (17)  $\calb(t-s)$ is a 4-point  self-correlation function
of the type $<v^xx(t-s)yy(t-s)>,$
gauging the degree of correlation in the equilibrium dynamics of the stars:
For a purely harmonic interaction potential 
of proper frequency $\omega_0/2,$
the function $\calb(t-s)$  was  
found to scale as $\sin[\omega_0(t-s)].$
Interestingly, we here notice that for such a potential the correlation time scale is
infinite. 
To illustrate this property, let us consider a more general expression  of the 
self-correlation function 
\begin{equation}
\calb(t-s)\propto \int_{-\infty}^{+\infty}d\omega \sin(\omega
(t-s))\exp\left
[-{(\omega-\omega_0)^2\over\sigma^2_{\omega}}\right ]
\end{equation}  
obtained weighting the sinusoidal function, that describes  the periodic
nature of the
orbits of the background particles, with a Gaussian centered about
$\omega_0$ with
dispersion $\sigma_{\omega}.$
We have included such a dispersion to mimic the 
characteristic spread in the frequencies of the background particles motions.
The integral can be evaluated straightforwardly
\begin{equation}
{\cal {B}}(t-s)\propto \sigma_{\omega} \sin(\omega_0(t-s)) \exp\left
[-\left ({\sigma_\omega (t-s)\over 2}\right)^2
\right ]
\end{equation}
to show that $\calb$ acquires an intrinsic cutoff time scale
\begin{equation}
\tau^*=2/\sigma_{\omega}
\end{equation} 
which is determined by the  dispersion ($\sigma_{\omega}$) in
the orbital frequencies
of the equilibrium 
system.  
If the distribution is sharply peaked about $\omega_0/2$ 
(as for the harmonic potential) $\tau^*$ is exceedingly large and  
the inclusion of  the actual 
dynamics is essential in determining the drag.
As illustrated by equation (20) $\tau^*$ is finite when more
frequencies are contributing to the dynamics in the virial spherical system.
Hence, the ``richness'' in the spectral decomposition of the orbits is an 
indirect measure of the self-correlation time, a quantity that has to be 
compared  with the typical radial  period time $T_r$ and with
the  characteristic time $T_{in}$ of interaction (of order of
$T_r$ for the cases explored in this paper) between the
satellite and the stars to determine the importance of the real
dynamics in affecting the drag.

\subsection {On Global Tides, the Wake and the Shift of the Barycenter}

A longstanding question is  whether dynamical friction in
self-gravitating backgrounds  is a local or global process.

As shown in C98, the satellite excites a tidal deformation when
orbiting outside the halo: This deformation is clearly global as it
involves the whole galaxy volume.  A global response is excited also
in shortlived flybies deep across the halo, giving a force along
$\vecV$ and $\vecR$ which is proportional to the background density
gradients (CP).  But in addition to such a global response, a back
reaction force rises due to the overdensity that the satellite excites
along its path (eq. [42] and [47] of CP).  This is usually referred to as
being the ``local"
contribution to the drag depending on intensive quantities like the
background density $n_0$, despite the presence of a Coulomb
logarithm that accounts  for those ``distant"
encounters which are effective for the satellite's drag.
These examples (CP) illustrate that
the global tidal field and the wake are aspects of the response
that are simultaneously present; they can be clearly distinguished  in
the high velocity limit.

When considering the interaction  along bound orbits inside the halo 
the two contributions 
are  technically difficult to separate out. 
Nonetheless an approximate estimate of
the degree of locality of the response can be inferred selecting from the force
$\F$ its component along $\vecV$ ($F_V$) and comparing it with  the
frictional
force
\begin{equation}
\vF_{\infty}=-4\pi [GM]^2\,mn_0\ln\Lambda\left (
{\rm{erf}}(x)-{2x\over \pi^{1/2}}{\rm {e}}^{-x^2}\right ){\vecV\over\vert
\vecV\vert}
\end{equation}
from an infinite  homogeneous non self-gravitating collisionless background
(Chandrasekhar
1943; Binney \& Tremaine 1987); $x$ is equal to $\vert
\vecV\vert/\sqrt{2}\sigma.$

Figure 7 shows that $F_V$ (filled
dots linked with solid 
line) is maximum just after each pericentric passage, the lag being a
manifestation of the memory effect. $F_V$ accounts for nearly $80-90\%$ of
the total force that also has  a component along $\vecR.$

The force $\vF_{\infty}$ (dashed line) is computed 
using the value of the density $n_0$ and
the dispersion velocity $\sigma$ at
the (``local") instantaneous 
satellite's position, and  
setting $\ln\Lambda=\ln(r_t/\epsilon)$ a value which is close  to
$\ln(Nm/M)
$. 
In $\vF_{\infty}$ the time lag is absent, and a closer analysis of the
two forces reveals that $\vF_{\infty}$ would predict a more rapid sinking 
than $\F.$ 
A time dependent Coulomb
logarithm can  better fit the
orbit and the evolution of the angular momentum, particularly in two
regions: at the periphery
where  
$\ln\Lambda\sim \ln
[(r_{apo}-R(t))/\epsilon]$, and close to the halo's center
where  $\ln\Lambda\sim
\ln[R(t)/\epsilon]$ (see the  dot-dashed line of Fig. 7).
In general we find  that it is difficult to reproduce accurately the
evolution  over a complete sample of
orbits using equation (21) since the component of the force along $\vecR$ 
gives a non negligible contribution: Related to the tides and
to the spatial inhomogeneities, this component varies in each single path.

Customarily $\ln\Lambda$  gives indication of the interval of
background particles impact parameters
for which the encounter is effective.
Can we have an intuitive understanding of the fits introduced above  ?
As guideline let us consider  
the temporal evolution of the density perturbation in a uniform background:
$\Delta n$ is found to result from  the composition of disturbances 
that originate at earlier times  $s$  
\begin{equation}
\Delta n(\vr,t)=\sqrt{ {2\over
\pi}}
GM n_0 \int_{t_0}^{t-\tau_{\epsilon}}
{ds\over \sigma^3 (t-s)^2} \exp[-{1\over 2}\Gamma_s(\vecR(s)-\vr)^2]
\end{equation}
which   are Gaussian in space, spherically symmetric  about 
$\vecR(s),$ and with  a characteristic length
\begin{equation}
\lambda_s={\Gamma_s}^{-1/2}= \sigma(t-s)
\end{equation}
(In equation (22) 
$\tau_{\epsilon}$ is introduced to mimic short distance
encounters yielding a finite minimum impact parameter.)
Since  the characteristic scale length becomes increasingly small
as  $s\to t-\tau_e$, the deformation  is large primarily in the vicinity
of the satellite where $\vert\vecR(s)-\vr\vert\ll \lambda_s$.
At earlier times $s\ll t-\tau_{\epsilon}$ the Gaussian 
disturbance   has a wider extension
indicating that the density perturbation broadens 
in space and weakens in magnitude being a transient structure.
(Only in the high speed limit (i.e., $\sigma/V\to 0$)
the overdensity develops in a sharp edge, a shock that never broadens
as stars behave as a cool continuum, i.e., as dust.)
In an infinite medium  the decay of the overdensity is not
sufficiently rapid to make the drag finite and this is the reason why
a cutoff distance, of the order of $r_t,$ is introduced in the
expression of $\ln\Lambda$
(eq. [21] is derived from eq. [6], and is a test on TLR).

In a spherical halo, the wake develops only as soon as  the satellite 
enters the stellar medium, so 
the maximum impact parameter varies with time initially, as suggested by
our first fitting formula.
Later, the wake  spatially widens while
fading across the  medium, in analogy with equation (22) at a rate 
which  increases
with decreasing  distance $r,$ being $\propto \sigma$ (eq. [12] and [22]).
In bound nonuniform  systems there is the tendency
of
erasing more rapidly the memory of the perturbation than in an infinite
medium, yielding to a weaker drag and to a force  
along $\vecV$ which is influenced more by the local
properties of
the background. This 
is likely a consequence of $\tau^*$ being smaller than the 
internal dynamical time.
Nonetheless, the actual dynamics of the satellite
can  be determined only within TLR which gives the description of  the
full stellar response (including tides and the effect of a nonuniform
background.

The coherent shift of the halo's center of mass is an important
aspect of the response
(White 1983; Weinberg 1989; Hernquist \& Weinberg 1989; Prugniel \& Combes
1992;
S\'eguin and Dupraz 1994, 1996; Cora et al. 1997).
Its inclusion accounts for the correct estimate of the global large scale density
deformations induced by the satellite; pinning the center of mass of the primary
(i.e., not including the shift) would  result
in more intense  tides that instead are not excited in a real encounter.  
We have verified that this correction
becomes important when the satellite is set on progressively wider orbits 
and of low eccentricity. 
For the vBLLS models the correction
on
the sinking
times accounts for about 10$\%.$
It is larger $>40\%$ when the satellite mass increases
(we explored a few cases with $M/Nm=0.08$ and $r_{cir}/r_{t}=1)$ and goes
always in the
direction of reducing the extent of the drag.
Weinberg included in his formalism the shift of the  barycenter
of the primary system 
coupling the Boltzmann equation for $\fop$ to the Poisson equation for the density
perturbation.  This approach makes the calculations too complex and does
not allow for a simple expansion of the drag force in powers of  the coupling
constant $G.$ 

\section {The Sinking of the Deformable Satellites or: How Does Tidal Stripping Affect Orbital Evolution ?}

In the previous sections we have carried out a detailed study of
dynamical friction using TLR and $N$-Body simulations to gain insight
into the physical mechanisms that cause the braking of a satellite and
its subsequent orbital decay.  In the cases explored in $\S4$ the satellite was
treated as rigid body while real satellites are deformable systems,
comprising a small luminous component hosted by a massive and extended
dark matter halo.

The outer part of the dark matter halo can be strongly damaged by the
tidal field of the primary, while experiencing dynamical friction.  As
a consequence, a reasonable picture of the evolution of satellites has
to take into account the role of tidal forces as well as dynamical
friction.  Here we determine how mass stripping affects evolution.

Among the satellites of the Milky Way, a few have experienced at least
one or two close pericenter passages and have suffered mass loss by
the global tidal field of the Milky Way: a clear example is
Saggitarius A which at present is in the verge of being disrupted
(Ibata \& Lewis 1998).  Using  $N$-Body simulations 
we have followed the evolution of satellites,
described as spherical halos, to explore the interplay between mass
stripping and orbital decay due to dynamical friction. 
We then tried to fit the numerical results
within the framework presented in the previous sections.

\subsection{Initial conditions}

Large cosmological $N$-Body simulations within the cold dark matter
(CDM) framework show that (satellite) halos have density profiles
which can be fit by so called NFW or Hernquist density profiles (Navarro et al.
1996, 1997; Ghigna et al. 1998). These profiles have a central cusp and
fall steeper than the isothermal profile at large radii. However, the
resulting rotation curves are in conflict with those observed for
dwarf galaxies and low surface brightness galaxies which exhibit a
large core in the center (Moore et al. 1999a; Persic \& Salucci
1997).  Feedback due to mass outflows of baryons as a consequence of
supernova-driven winds have been invoked to reconcile this discrepancy
(Navarro et al. 1996; Gelato \& Larsen 1999) but the solution of
this problem still awaits (Burkert \& Silk 1999).
We thus employ truncated isothermal profiles wit cores to model satellite
halos, as these allow good fits with observed rotation curves 
for different galaxies (de Blok \& McGaugh 1997).
The primary galaxy is represented by the same model used in the
simulations described in $\S4$.

We build three different models for the satellite: these 
have the same virial mass $M$, which is  $0.02Nm,$ but differ in the
value of the concentration  $c,$  where $c$ is the ratio
between the satellite's tidal radius $r_t^s$ and its core radius
$r^s_c.$  
This parameter sets the
value of the
central density as  $\rho_{0} \propto c^{2}.$
Tidal damage of satellites' halos should be basically related
to the ratio between their own central density and that of
the primary halo: for this reason the concentration of the
satellite's density profile cloud play an important role in deciding
its final fate.
A general result of hierarchical clustering is that lower
mass halos are on average denser because they formed earlier,
when the background density of the Universe was higher.
An analysis of cosmological simulations shows that
the characteristic
halo density scales as ${M}^{-\nu}$
(Syer \& White 1998). The value of
$\nu$ is
related to the slope of the power spectrum on the scale of interest and
is $\simeq 0.33$ for galaxy-sized halos in a standard CDM
cosmogony: according to this estimate an LMC-like satellite
should have a central density $\sim 4$ times higher than that
of the Milky Way. 

The reference model (S1) for our dark matter satellite is simply
a rescaled  version of the primary
galaxy, according to the relations $r^s_t/r_t=(M/Nm)^{1/3}$ and
$V^s_{cir}/V_{cir} = (M/Nm)^{1/3}$ (White \& Frenk 1991). 
%%where  $V^s_{cir}$ and $V_{cir}$ are their circular velocities. 
The resulting satellite  has a circular 
velocity $V^s_{cir}$ of about $50$ km s$^{-1}$, very close to
that of LMC.
The other two models have a concentration which is
two times (model S2) and three times (model S3) that of the reference 
model S1.
We use 10,000 particles for the satellite models. One simulation
has been rerun with 50,000 particles as a test, giving practically
identical results.
We employ the
same system of units as in vBLLS, along with the same timestep
and softening for the primary system. The softening for the  satellite  
scales as $(M/Nm)^{1/3}$ relative to  the primary.
We have considered only orbits with $e=0.8$ and $e=0.6$ for both
$r_{cir}/r_t=1$ and $0.5$.
Models S2 and S3 have been run only for the most destructive encounter,
i.e. for $r_{cir}/r_t=0.5$ and $e=0.8$.

\subsection {The fate of galaxy satellites}

Our results allow for a clear interpretation of the interplay between
dynamical friction and mass loss due to the tidal field of
the primary.
Satellites lose on average about $60 \%$ of their mass after
the first pericentric passage (at 1.5 Gyr), while  their orbital angular
momentum has decreased by no more than $20\%$ (as illustrated in Fig. 8):
this means that tidal stripping is always more efficient than
dynamical friction.
Their final fate  however depends on their 
initial concentration, and, though less sensitively, on the  orbital
parameters. Model S1 is  disrupted over a time comparable to the 
dynamical friction decay time $\tau_{DF}$ of its rigid counterpart,
just after the second pericenter passage.  
(In a test with a $5 \times 10^4$ particles on a $r_{cir}/r_t=0.5$ and $e=0.8$ 
orbit, the satellite (model S1) was disrupted nearly at the same
time as in the runs with $10^4$ particles. This proves that the 
resolution used does not affect significantly the physical
interpretation of our results.)
Satellites with a high density contrast (S2 and S3) relative
to the primary central density can instead survive even the third
pericenter passage (at $5$ Gyr)
despite being on eccentric tightly bound orbits. They will then suffer
disruption along their orbit, after $6-8$ Gyr (performing a total of $4-5$
pericenter passages).
One has to bear in mind that these high-concentration satellites more
likely correspond to the halos of small galaxies found in cosmological CDM
simulations. The case of model S1 could instead represent the
halo of low surface brightness  satellites that typically have  large
cores and 
still have to find an explanation in a cosmological context.
All satellite halos survive much longer than $10$ Gyr, regardless of their
concentration  if they move on the peripheral orbits (with
$r_{cir}/r_t=1$).
In these cases dynamical friction almost
switches off, because of mass loss, when the satellite is still far from
the densest region of the primary.
As for the rigid case, orbital decay is not accompanied by significant
circularization.

The reduced effectiveness of dynamical friction as a result of mass
loss has important consequences for the merging of the baryonic
components inhabiting dark matter halos. 
In principle, the loss of orbital angular momentum implies a decrease in the
sinking time, depending on $r^2_{cir}$. On the other hand the mass loss
implies an increase in the sinking time, which scales  as $M^{-1}$. The 
simulations have shown that substantial  mass loss occurs already at the
first pericentric passage, when the angular momentum has not significantly
decreased yet.
The satellite orbit has thus not decayed sufficiently 
(i.e. $r_{cir}$ has only slightly decreased) to
counterbalance the reduction in mass and
the overall result is that the orbital decay will be considerably slowed
down.  
The central region would survive the subsequent 
disruption of the outer  dark matter
halo, being more compact (Mayer et al. 1999; Navarro et al. 1995, 
Ghigna et al. 1998) and would then decay on a very long timescale.
A decoupled orbital evolution of dark and baryonic
components was already suggested by Lacey \& Cole (1993) in
their semianalytical treatment of galaxy merging rates. In that case it
was implicitly assumed that satellites lose immediately their dark matter
halos, finding themselves on a bound orbit at the periphery of the primary
halo:
the dynamical friction time was then computed with formula (15) adopting
merely the baryonic mass for the mass of the satellite.
Navarro, Frank \& White (1995) instead were led to suggest that 
satellite galaxies merge at the center of the primary halo on a time scale
determined by their initial total mass (baryonic + dark): they came
up to this conclusion by directly comparing the prediction of formula (15)
with the sinking times of gaseous cores in $N$-Body/SPH simulations of
galaxy formation.
Their distribution of merging times showed however a large scatter with
respect to the analytical prediction: moreover, a significant group of
satellites existed with sinking times $2-3$ times larger than the analytical
estimate.
This turned out to be satellites with an initial mass $< 1/10$ of their
primary halo, i.e. they were in the mass range of typical galaxy satellites.
Our results suggest the higher efficiency of tidal stripping with respect
to dynamical friction is responsible for such an increase of merging times of
galaxy satellites.

We can now give an estimate of the merging time that incorporates mass
loss as well as orbital decay.
We can approximate the mass loss curve obtained
for satellites moving on eccentric tightly bound orbits
(which are the most likely for satellites) with an exponentially
decreasing function of time given by:
\begin{equation}
M(t)=M_{d}\exp(-t/T_r) + M_{b}
\end{equation}
where $M_{d}$ is the dark mass of the satellite, corresponding roughly to its
initial total mass, $M_{b}$ is the baryonic mass, which we assume to
be $< 1/10$ of the total mass, and $T_r$ is the orbital radial period. 
The satellites continue to lose mass at
every pericenter passage: however, most of their mass is stripped already
on the first orbit and this occurs independently of their concentration.
Moreover, at this time the orbital parameters are still very close
to the initial ones due to the low efficiency of the orbital decay. 
We then use the tidally limited mass at the first pericentric
passage $(M_{d}/\rm {e})$ as the  ``effective" mass for the satellite in
formula
(15).
% as well as adopting the initial orbital parameters.
The merging time of the satellite galaxies $\tau_{m}$ can be then computed
using 
\begin{equation}
\tau_{m}=1.2{
J_{cir}r_{cir}
\over[GM_d/{\rm{e}}]\ln(M_{halo}/M_{d})}\varepsilon^{0.4} 
\end{equation}
where $J_{cir}$ and $r_{cir}$ are, respectively, the initial
orbital angular momentum and radius of the circular orbit with
the same energy of the orbit on which the satellite is placed
This formula updates the one by Lacey \& Cole incorporating both
the different normalization factor and eccentricity dependence
as well as the ``delaying'' effect due to tidal stripping.
The reliability of this prescription was tested by running a simulation
with a rigid satellite on an orbit $e=0.8$ with an initial mass reduced
by a factor $(1/\rm{e})$ with respect to the standard mass and then
comparing
the angular momentum loss in this case with that occurring in the
corresponding run with the deformable satellite (Fig. 9).
These are remarkably close, while much more angular momentum is lost by 
the rigid satellite with the standard mass, which suffers complete orbital
decay (Fig. 9).
Our estimate suggest that the satellite galaxies would merge on a timescale 
almost $2-3$ times larger than previously estimated with (15). In
addition, our $N$-Body simulations indicate that the survival time $t_{s}$
of the dark matter halos of the satellites falls between $\tau_{DF}$ and
$\tau_{m}$, with more concentrated haloes surviving for a longer time. 
However, the presence of a baryonic core inside the halo can prolong
the lifetime of the central part of the halo itself 
because the overall potential well becomes deeper
(Mayer et al. 1999).
Only very large satellites (i.e. satellites with a mass $>1/10$ relative
to that of the primary) could decay on a short timescale compared
to the Hubble time, as the rate of orbital angular momentum loss
would be comparable to the rate of mass loss (this being related
only to the ratio of the central densities of the two systems and not
to their masses).
Such events could have occurred in the building up of our
galaxy, at $z > 1$ when the progenitor halo at that time probably
accreted big lumps on orbits with small pericenters  
(because the virial radius of the primary was smaller), a
condition which should further reduce dynamical friction time scale (see
also Kravtsov \& Klypin 1998).
Satellites accreted after that epoch had not enough time to 
decay and merge and these are the present-day satellites of the
Milky Way and Andromeda.

\section {Discussion and Conclusion}

In this paper we present, for the first time, a unified view of the
physical process responsible for the braking of a satellite in a
self-gravitating spherical stellar system.  We have shown that TLR
embraces all aspects of the gravitational interaction of a massive
object with a background of lighter (dark matter) self-gravitating
particles.

We have found that the characteristic dynamical friction time for
satellites with mass ratios $<0.1$ (moving on cosmologically relevant
orbits) in a galaxy like our Milky Way is sufficiently long that
satellites have not merged with the stellar disk yet.  As first
noticed by Ghigna et al. (1998), orbital decay is not
followed by a significant circularization when the satellite happens
to orbit well inside the primary halo.  The expected equilibrium
distribution of eccentricities in a spherical potential is skewed
toward high $e\simeq 0.6-0.7$ (see vBLLS) and dynamical friction plays
no role in modifying such a distribution.  We have in addition shown
that rigid satellites on eccentric orbits have sinking times only
slightly shorter than those of satellites on circular orbits: previous
analysis instead predicted a wider spread, implying a much shorter
lifetime in the primary halo.  Tidal stripping is more efficient than
dynamical friction for the typical masses of galaxy satellites. We
have shown that this should considerably prolong the lifetime of the
baryonic lumps inhabiting satellite halos with masses up to $0.1$ that
of the primary halo, so that they will wander along their orbit for at
least an Hubble time.  This has important implications for many issues
concerning the evolution of galaxies.  We indeed expect that those
satellites that entered the primary halo after $z\sim 1$ cannot have
decayed to the center yet, while their dark matter halo has been
already substantially stripped.  These correspond to the present-day
satellites of spiral galaxies, like those populating our Local Group.

Our results show that the disk is unlikely to have suffered any late
merging event or penetrating encounter with a typical 
satellite.  The overall picture which emerges is that
substructure survives longer than previously believed and this nicely
agrees with numerical findings in large cosmological simulations
(Ghigna et al. 1998; Tormen et al. 1998; Klypin et al. 1999).  In general,
the
pericenters of the orbits of satellites are reduced by no more than a
factor of $2$ in about $7$ Gyr, which is approximately the time which
passes between $z=1$ and $z=0$.  This large population of almost
indestructible satellites could have a dramatic effect on the dynamics
of spiral galaxies disks. Work by Moore et al. (1999b)
suggests that the cumulative effect of many nearby encounters between
this numerous population of small satellites on almost radial orbits
and a galactic disk would heat its stellar component considerably on a
time scale of a few Gyrs even without merging with it.

Dynamical friction and tidal stripping are among the main dynamical
mechanisms involved during the formation of cosmic structures. Our
results provide a detailed description of these processes as well as
giving the theoretical support for understanding much of the
underlying physics.  Numerical simulations and theoretical models are
now converging towards a common picture where CDM models create a
large wealth of long lived substructure in dark matter halos.  It now
remains to investigate its effects and its observational evidence, so
providing tight constraints on theories of galaxy formation and
evolution.

\vskip 40pt
We thank J. Binney for enlightening discussion 
and T. Quinn and J. Stadel for kindly providing the PKDGRAV code.
This work acknowledges financial support from MURST.

\def \vF  {{\bf {F}}}
\def \F {\vF_{\Delta}}
\def \vri {{\bf {r}}_i}
\def \vr  {{\bf {r}}}
\def \vpi {{\bf {p}}_i}
\def \vp {{\bf {p}}}
\def \vvi {{\bf {v}}_i}
\def \vv {{\bf {v}}}
\def \vx {{\bf {x}}}
\def \vecX {{\bf {X}}}
\def \fit { {\tilde{\phi}} }
\def \er { {\rm{erf}} } 
\def \vw  {\bf {w}}
\def \grad {{\bf {\nabla}}}
\def \vecR  {{\bf {R}}}
\def \vecV  {{\bf {V}}}
\def \fop {{f^{\rm {op}}}}
\def \mr {{<r^2>^{1/2}}}
\def \vir {{<r^2>}}
\def \si {{<v^2>^{1/2}}}
\def\ltord{\hbox{$\;\raise.4ex\hbox{$<$}\kern-.75em\lower.7ex\hbox{$\sim$}
                           \;$}}
\def\gtord{\hbox{$\;\raise.4ex\hbox{$>$}\kern-.75em\lower.7ex\hbox{$\sim$}
         \;$}}

\vskip 40pt

\centerline {APPENDIX}

In this Appendix we shortly describe the semianalytical calculation
for the drag force $\F(t)$ acting on the satellite at time $t$  (eq. [6]).
The Force is computed under the hypothesis expressed in equation (8) 
considering a Gaussian distribution function 
in the velocity space for which
$\grad_{\vv}\fop=-(\vv/\sigma^2)\fop.$
Due to the complexity of the various expressions we introduce
the vector
$$
\vecX\equiv\vecR(t)-\vr,
\eqno(A1)$$
and  define     
$$
\grad_{\vecR(s)}\fit\equiv\grad_{\vecR(s)}\phi(\vert\vecR(s)
-\vr\vert)-\int 
d_3\vr'\,n_0(r')\grad_{\vecR(s)}\phi(\vert\vecR(s)-\vr'\vert).
\eqno(A2)$$
The integral on the velocity space can be written in a simplified form,
so  
the force reads
$$
F^b_{\Delta}(t)=-[GM]^2Nm\left( {1\over 2\pi}\right)^{3/2}
\int_{t_0}^t\,
{ds\over (t-s)^4}
\,\int d_3\vr {n_0(r)\over \sigma^5(r)}{\partial \fit\over \partial
R^a(s)} 
{\partial I^a\over\partial R^b(t)}
\eqno(A3)$$
where the vector   $\bf{I}$ is found to have components
$$I^a={2\pi\over \Gamma}{X^a\over \vecX}\int_0^{+\infty}\, dy \, y\,
\phi(y)\,\,\,\,\times
$$ 

$$\left \{
(X^2-Xy+\Gamma^{-1})\exp[-{1\over 2}\Gamma(X-y)^2]-
(X^2+Xy+\Gamma^{-1})\exp[-{1\over
2}\Gamma(X+y)^2]  
\right \}\eqno(A4)$$
with $\Gamma\equiv[\sigma(t-s)]^{-2}.$
Since $\bf {I}$ can be expressed  as gradient of a scalar function
we are able to determine the force as 
$$F^b_{\Delta}(t)=
[GM]^2Nm\left ({1\over 2\pi }\right )^{1/2}\int_{t_0}^t\,ds
\int d_3\vr\, {n_0(r)\over \sigma(r)} {\partial \fit
\over \partial R^a(s)}\,\,\,\times$$

$$ {\partial^2\over \partial R^b(t)\partial R^a(t)}{1\over
\vert \vecR(t)-\vr\vert}\int_0^{+\infty}\,dy\, y\, \phi(y)
\left \{\exp[-{1\over 2}\Gamma(X-y)^2]
-\exp[-{1\over 2}\Gamma(X+y)^2]                   
\right \}
\eqno(A5)$$
Equation (A5) involves integrals over the physical volume
of the halo, over  time $s$ and $y$.
The integral in the $y$ variable can be computed analytically, given the
expression
of the potential $\phi(y)$.
According to our analysis of $\S 3$, the function $\phi$ is 
just the spline kernel potential $\phi_{sp}$ introduced  by Hernquist \&
Katz 
(1989; we refer to his Appendix)
that reduces  to the Newtonian potential ($\phi=-1/y$) at distances
larger than
$2\epsilon$:
this potential is written as an expansion in powers of  $y.$ 
In equation (A5) we thus need to calculate integrals of the form
$$
\int_{inf}^{sup}\, dy y^n
\left \{\exp[-{1\over 2}\Gamma(X-y)^2]
-\exp[-{1\over 2}\Gamma(X+y)^2]
\right \}
\eqno (A6)$$
with $n\ge 0;$  the domain of integration $(inf,sup)$ 
is uniquely defined by the form of  $\phi_{sp}.$
If we introduce the functions 
$$B_n(a,\Gamma)\equiv B_n(a,X,\Gamma)-B_n(a,-X,\Gamma)\eqno(A7)$$
 with
$$B_n(a,X,\Gamma)\equiv \int_0^a\,dy y^n  \exp[-{1\over
2}\Gamma(X+y)^2]
\eqno(A8)$$ 
and 
$$B_n(a,-X,\Gamma)\equiv \int_0^a\,dy y^n  \exp[-{1\over
2}\Gamma(y-X)^2]
\eqno(A9)$$ 
we can express $B_n(a,\Gamma)$ as a linear combination of Error 
Functions, and the following recurrence relations apply:

$$B_0(a,X,\Gamma)=\left({\pi\over
2\Gamma}\right)^{1/2}\left \{ \er[\sqrt{(\Gamma/2)}
(X+a)]-\er[\sqrt{(\Gamma/2)}X]\right \}\eqno(A10)$$

$$B_1(a,X,\Gamma)=-XB_0(a,X,\Gamma)-{1\over \Gamma}
\left \{\exp[-{\Gamma\over 2}(X+a)^2]-\exp[-{\Gamma\over 2}X^2]
\right \}
\eqno(A11)$$

$$B_{n+1}(a,X,\Gamma)=
-XB_n(a,X,\Gamma)+{n\over  \Gamma}B_{n-1}(a,X,\Gamma)-{a^n\over\Gamma}
\exp[-{\Gamma\over 2}(X+a)^2].\eqno(A12)$$
The integral on the $y$ variable in  equation (A5) is constructed using
equations (A6-A12).

Given the above relations we can calculate also the first  
and second 
derivatives of $B$, $B'$ and $B'',$  relative to $X;$ after a
number of
simple but long
steps we can
express the drag force as
$$
F^b_{\Delta}=-[GM]^2Nm\left ({1\over 2\pi}\right )^{1/2}
\int d_3 \vr {n_0(r)\over \sigma} 
\times$$

$$\int \,ds \nabla^a_{\vecR(s)}\fit\, \left \{
\left [{3 X^aX^b\over X^5}-{\delta^{ab}\over X^3}\right ]
\sum_n c_n \left [ B_n([d],\Gamma)-XB'_n([d],\Gamma)\right ]
+{X^aX^b\over X^3}\sum_n c_n B''_n([d],\Gamma)\right \}
\eqno(A13)$$ 
where  $B_n([d],\Gamma)$
shortly denotes that the functions of equations (A6-A9) are computed 
over the whole domain $(0,\infty)$ which is 
divided into three parts $(0,\epsilon)$, $(\epsilon,2\epsilon)$
and $(2\epsilon,+\infty)$ (see Hernquist \& Katz 1989).   
The coefficient $c_n$ in equation (A13) contains the $n$th power of
the softening length $\epsilon,$ as shown in the expression of  
$\phi_{sp}.$
The time and spatial integrals are computed numerically,
using standard  procedures, given the density
and velocity dispersion profiles (eq. [9] and [12]).

\clearpage

%bo.ps
%\figcaption{
\begin{figure}
\centering
\psfig{file=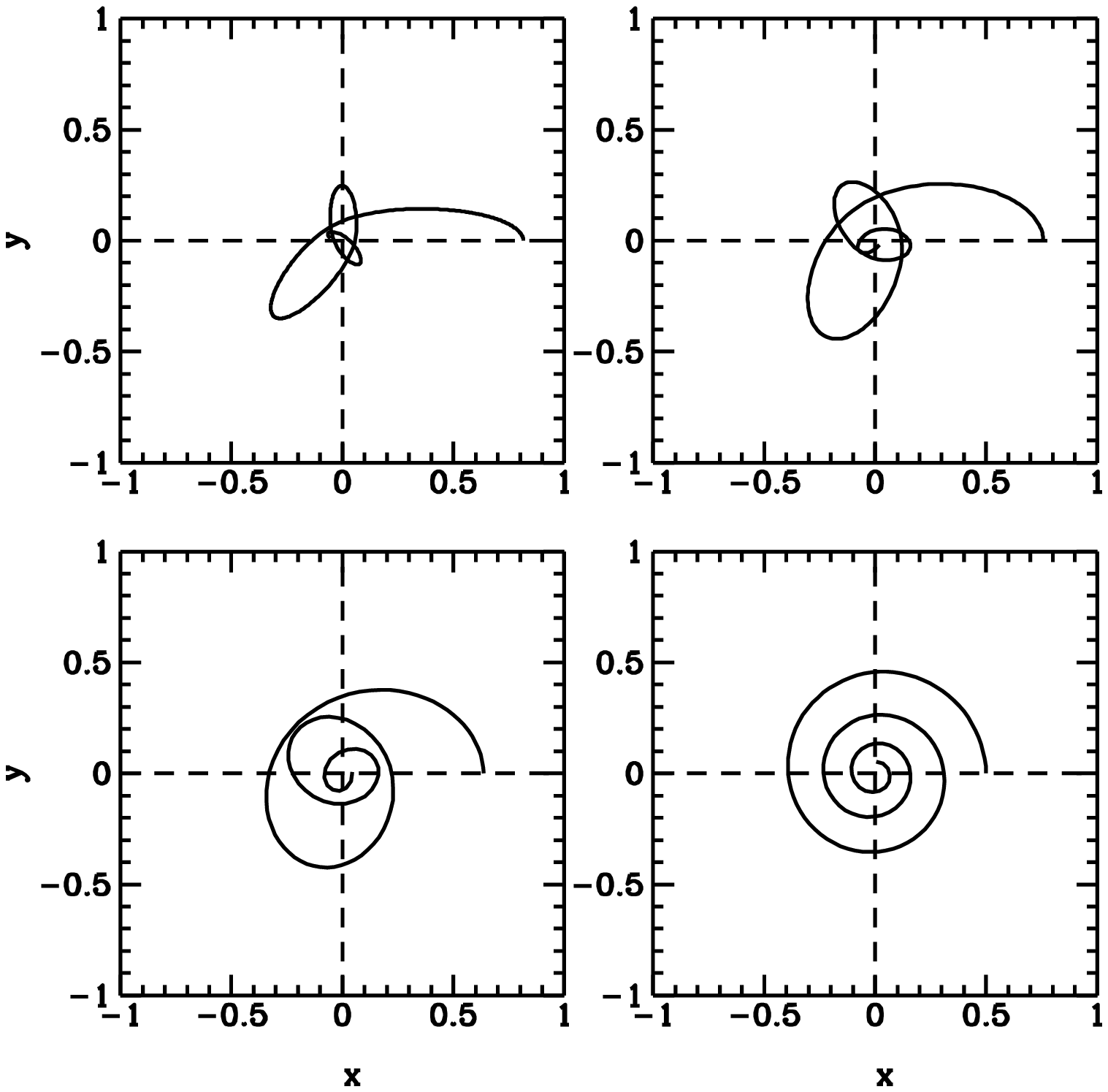,width=15cm}
\caption
{
Collection of orbits in the plane $(x,y)$ computed within TLR,  
for $r_{cir}/r_t=0.5$. From top left
to bottom right the initial eccentricity is  $e=0.8,0.6,0.3,0.$
respectively. Length is in units of $r_t.$}
\end{figure}

% bj.ps 
%\figcaption{
\begin{figure}
\centering
\psfig{file=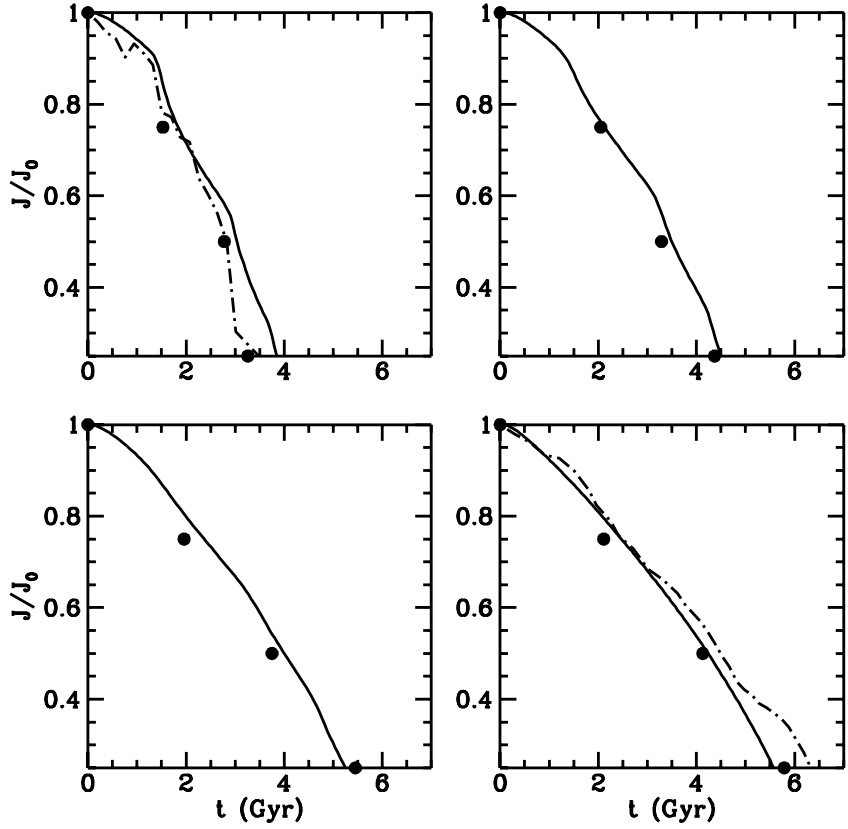,width=15cm}
\caption
{Orbital angular momentum $J$ (in units of the initial value $J_0$) as a
function of time for $r_{cir}/r_t=0.5$. From top left
to bottom right the eccentricity $e$ is $0.8,0.6,0.3,0.$ 
Solid lines indicate 
the results of TLR, filled circles  are the results of  vBLLS, and
dot-dashed lines are the results 
of  our $N$-Body runs.}
\end{figure}

%b1fin.ps
%\figcaption{
\begin{figure}
\centering
\psfig{file=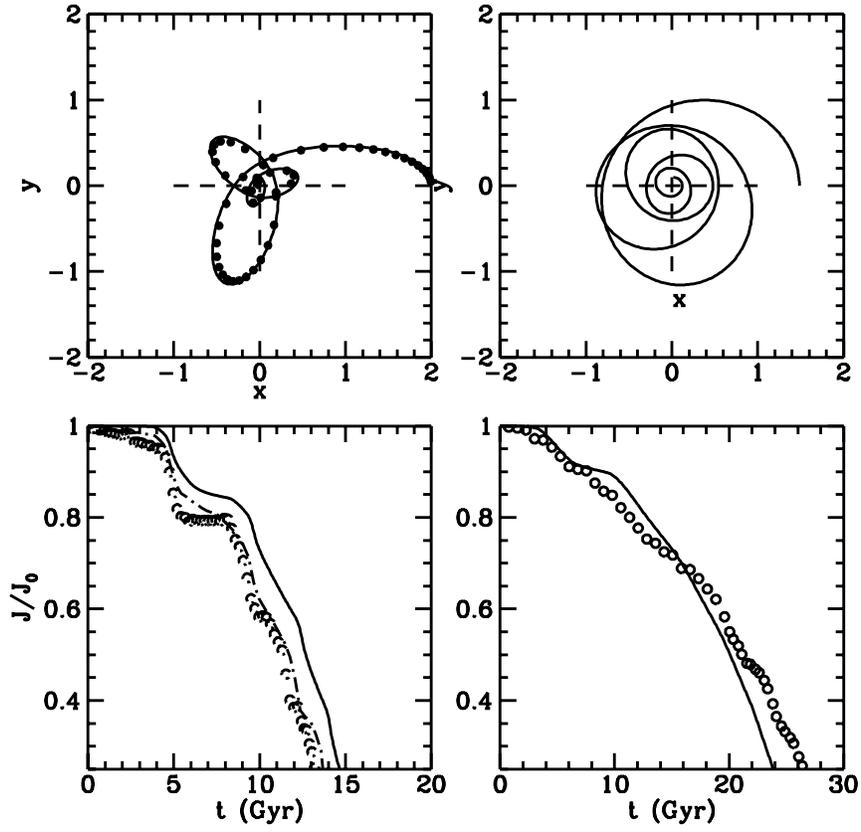,width=15cm}
\caption
{In the upper panel we show the satellite's orbital evolution 
computed within TLR (solid line) for $r_{cir}/r_t=1 $ for $e=0.8$
(left) 
and
$e=0.3$ (right). Filled circles are from our $N$-Body simulation using
100,000 particles.
In the lower panel we report the corresponding evolution of $J/J_0$.
The dot-dashed line in the lower left panel is the result of
the simulation with 100,000 particles. 
Open dots are from the $N$-Body runs with 20,000 particles.
Length are in units of $r_t$, and time is in Gyr corresponding to
1.34 time units.}
\end{figure}

%lacey.ps
%taudf.ps
%\figcaption{
\begin{figure}
\centering
\psfig{file=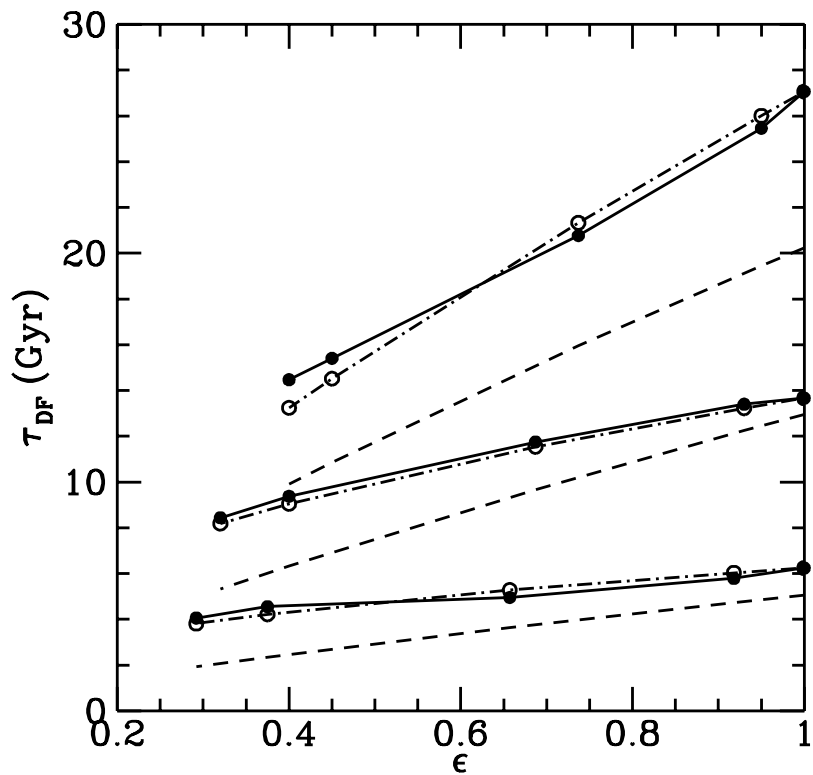,width=15cm}
\caption
{Dynamical friction time scale $\tau_{DF}$ (in Gyr) against circularity
$\varepsilon=J/J_{cir}.$ 
$\tau_{DF}$ refers to the time at which  $J/J_0=0.01.$
Filled Dots connected
with solid line denote the decay
times computed within TLR, for $r_{cir}/r_t=0.5$ (bottom curve),
$0.8$ (mid curve) and $1$ (top curve). Open circles  (connected with
dot-dashed line) are the
fit with $\alpha=0.4$ (bottom), $\alpha=0.45$ (mid) and $\alpha=0.78$
(top). 
The dashed line  denotes the fitting formula by Lacey \& Cole, given
by eq. [15].
}
\end{figure}

%eccen.ps
\begin{figure}
\centering
\psfig{file=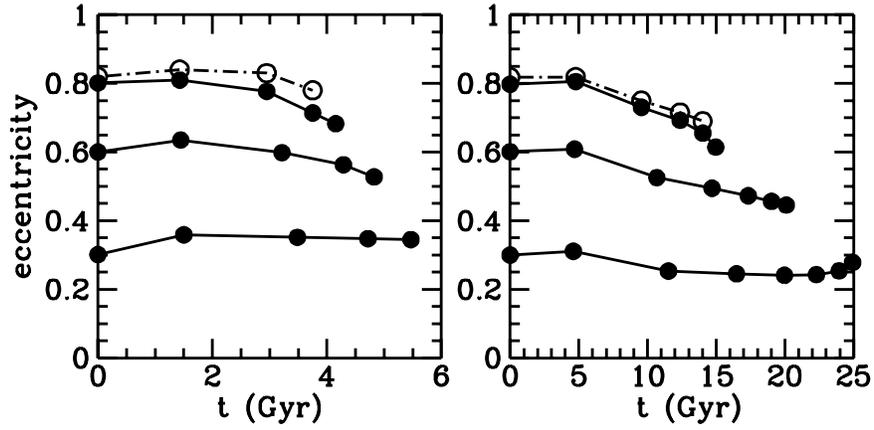,width=15cm}
\caption
{Eccentricity as a function of time $t$, for $r_{cir}/r_t$ equal to $0.5$
(left  panel)
and $1$ (right panel), and initial $e=0.8,0.6,0.3$.
Filled dots are from TLR while the open dots are from the $N$-Body runs.
Eccentricity is computed at each pericentric passage considering 
 a complete cycle along the actual
satellite's orbit. 
}
\end{figure}

%mem.ps
\begin{figure}
\centering
\psfig{file=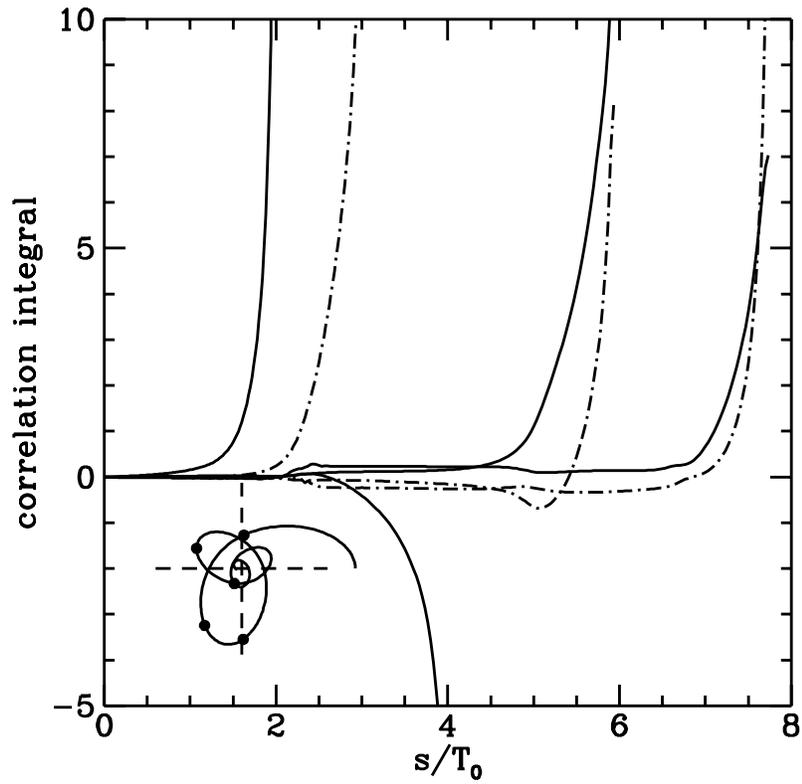,width=15cm}
%\figcaption{
\caption
{The correlation integrals  $I^x (s)$ (solid line) and $I^y (s)$ (dot-dashed line)
as defined in eq. [16] with $s$ varying from $t_0=0$ up to $t$
with time in dimensionless units.
A view of the orbit (with $r_{cir}/r_t=0.8$ and $e=0.6$) in the $(x,y)$
plane is drawn, where dots denote the
times $t/T_0$ equal to 2,3,4,6,8 at which
the functions $I$ (proportional to the components of the drag force and expressed in
dimensionless units) are computed.}
\end{figure}

%ln.ps
\begin{figure}
\centering
\psfig{file=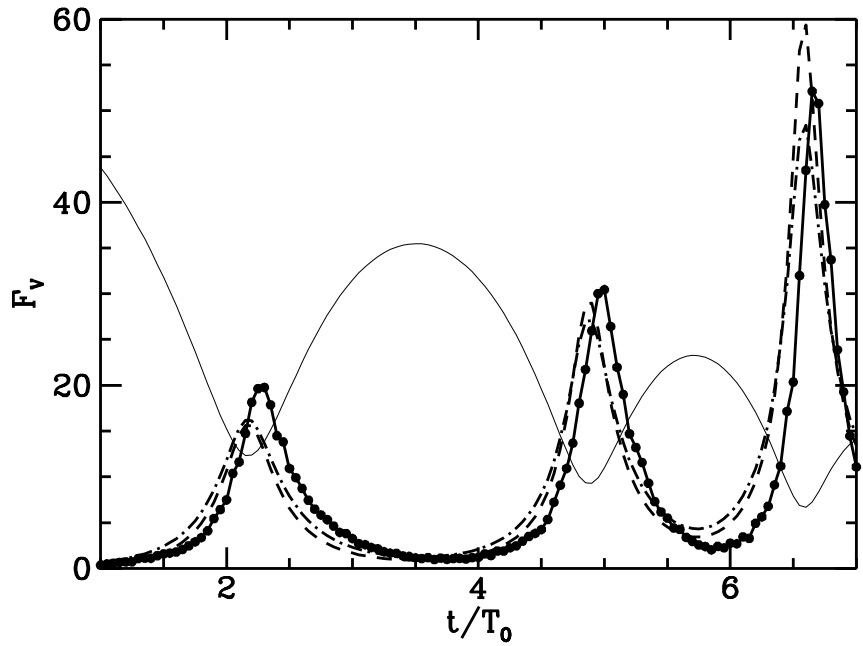,width=15cm}
%\figcaption
\caption 
{Filled circles connected with solid line denote 
the component of $\F$ along $\vecV$ (in arbitrary units) as a function 
of the dimensionless time. 
The dashed line is ${\bf {F}}_{\infty}$ evaluated for
$\ln\Lambda=\ln(r_t/\epsilon)$ 
while the
dot-dashed line gives ${\bf {F}}_{\infty}$ for
$\ln\Lambda=\ln(R(t)/\epsilon)$. In the background (
lighter solid line on an
arbitrary scale) 
we have drawn the current satellite distance $R(t)$ relative to
the stellar center of mass.}
\end{figure}

%jmstrip.ps
%\figcaption{
\begin{figure}
\centering
\psfig{file=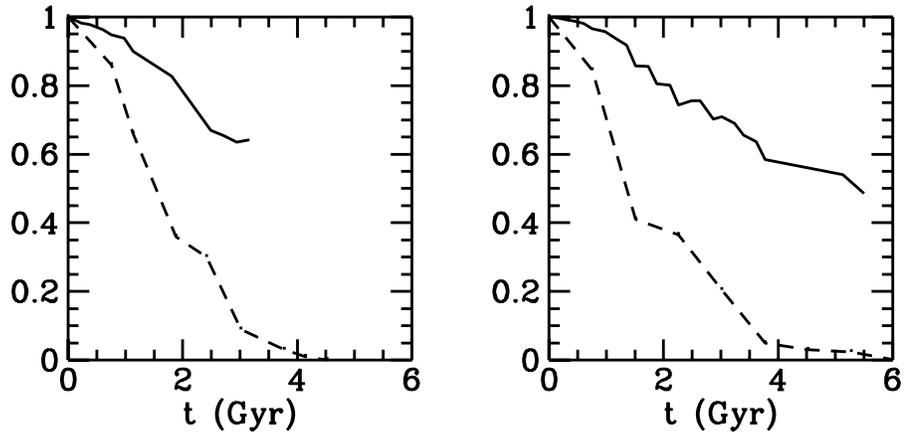,width=15cm}
\caption 
{Comparison between mass loss (dashed line) and orbital angular
momentum loss (solid line) as a function of time, for
model S1 and $r_{cir}/r_t=0.5$.
The left panel refers to 
an orbit with  $e=0.8$, the right 
is for $e=0.6$}
\end{figure}

\begin{figure}
\centering
\psfig{file=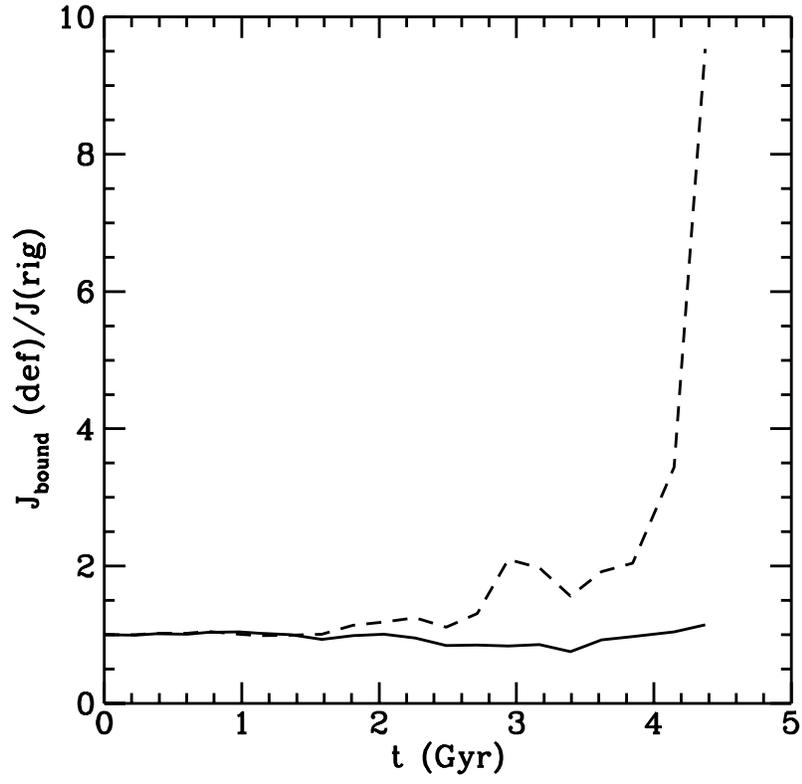,width=15cm}
\caption
%\figcaption
{ Ratio of $J(t)/J_0$ for the bound matter of the deformable S2 satellite
with $J(t)/J_0$ for the rigid satellite.
Curves are for  $r_{cir}=0.5$ and eccentricity equal to $0.8.$ Dashed line refers to a
rigid
satellite with
$M$ equal to the initial mass of the deformable satellite. Solid line refers to
a rigid satellite with mass reduced by the $\rm {e}$-folding  factor
$(1/\rm {e}).$ 
}
\end{figure}

\end{document}